\documentclass[12pt]{article}
\newcommand{\no}{\nonumber\\}
\newcommand{\be}{\begin{equation}}
\newcommand{\ee}{\end{equation}}
\newcommand{\ba}{\begin{eqnarray}}
\newcommand{\ea}{\end{eqnarray}}
\newcommand{\la}[1]{\label{#1}}

\def\gl#1{(\ref{#1})}

\def\Tr#1{\mbox{\rm Tr}\left[#1\right]}

\date{}

\begin{document}
\begin{center}
{\Large\bf
Dual oscillators and Quantum Pendulums: spectrum and correlators}\\
\bigskip

\bigskip

{\large A. A. Andrianov\footnote{E-mail: andrianov@bo.infn.it}}\\

{\it V.A. Fock Department of Theoretical Physics, Saint Petersburg State University, 198504, S.-Petersburg, Russia} \\
{\it Departament d'ECM, Universitat de Barcelona, 08028 Barcelona, Spain}\\
\end{center}
\begin{abstract}
We outline the nonlinear transformation  in the path integral representation for partition function of $O(N)$ symmetric oscillator systems bringing their duality to certain one-dimensional oscillators with unstable potential shapes. This duality transformation realizes the equivalence between spectra of a Hermitian and a non-Hermitian Hamiltonians. It is exploited to calculate exactly the spectra of quantum rotators . The zero-temperature limit is considered and the dual representation of n-point correlators of coordinate operators is obtained.
\end{abstract}

\section{Introduction}
The method of equivalent oscillators \mbox{(EO-transformation)} elaborated in \cite{AAA} allows one to find the correspondence between quantum oscillator systems with attractive and repulsive anharmonic interactions. This equivalence can be realized when the coordinate paths for the repulsive interaction are taken along a curve in the complex space \cite{AAA,ptsym} . Recently this kind of relationship between Hermitian and non-Hermitian oscillator systems has attracted much attention \cite{jones,bender} in the search for examples of PT-symmetric non-Hermitian quantum Hamiltonians with real spectra \cite{ptsym,complex4,dorey,mosta}.

In our paper we  present the derivation of EO-transformation using the phase-space-like path integral, find  a novel, general representation for dual QM system and apply it to the partition function of anharmonic oscillators and of quantum pendulum in order to calculate their spectrum by path-integral means. The $n$-point coordinate correlators are examined and their dual analogs are obtained.

 The basic EO-transformation evaluates the partition function ${\cal Z} (\beta),\ \beta = 1/kT$ for a $N$-dimensional harmonic oscillator with variable complex frequency $\tilde\omega^2 (\tau) = \omega^2  + i \sigma(\tau)$ assumed to be periodic, $\tilde\omega^2 (0) = \tilde\omega^2 (\beta)$ ,
\ba &&{\cal Z}_\beta(\sigma) = \Tr{\mbox{\rm T}\cdot\exp\Big(- \int\limits_{0}^{\beta} d\tau\hat H \big(\sigma( \tau)\big)\Big)}\no && = \int\limits_{\vec{q}(0) = \vec{q}(\beta)} {\cal D}^N q(\tau) \no&&\times\exp \left\{ - \int\limits_{0}^{\beta} d\tau \left[{\dot{\vec q}\ }^2(\tau) + \tilde\omega^2 (\tau) {\vec q \ }^2(\tau)\right]\right\} , \la{genf} \ea where $(\vec q) = (q_1,\ldots,q_N) $.  From the structure of the action in \gl{genf} it is seen that the imaginary part of the frequency may serve to realize the Legendre transformation to the description dual to the $O(N)$ invariant coordinate ${\vec q \ }^2$. The real part of the frequency $\tilde\omega^2 (\tau)$ could also depend on Euclidean time but in this case we assume its positivity $\forall\tau, \ \mbox{Re}\ \tilde\omega^2 (\tau) =  \omega^2 + \nu(\tau) > 0 .$ Let's skip the variable part $\nu(\tau)$  for a while as it can be always embedded by shifting $\sigma \rightarrow \sigma -i \nu$. Formally, after path integration one comes to the functional determinant of the differential operator $-\partial_\tau^2 + \tilde\omega^2$, 
\ba {\cal Z}_\beta(\sigma)  &=& C \parallel -\partial_\tau^2 + \tilde\omega^2 \parallel^{-N/2}\no &=& C \exp\{\Tr {\log(-\partial_\tau^2 + \tilde\omega^2 )}\} , \la{nonloc} \ea 
where the  constant $C$ is chosen so that for the constant $\tilde\omega = \omega$ the partition function is saturated by the discrete spectrum of the harmonic oscillator, \be {\cal Z}_\beta(0) = \left(2\sinh \frac{\omega\beta}{2}\right)^{- N} . \la{harm} \ee This analytic expression is not available when the function $\sigma(\tau)$ is variable in time Instead it is given by the functional nonlocal in $\sigma(\tau)$,
\ba &&{\cal Z}_\beta(\sigma) = \left(2\sinh \frac{\omega\beta}{2}\right)^{- N} \no &&\times\exp\Tr{ \log\left(I +\frac{1}{-\partial_\tau^2 + \omega^2}\ i \sigma(\hat\tau)\right)}. \la{varharm} \ea
The temperature Green function $G_\beta(\tau)$ is defined to fulfill the periodicity in Euclidean time $G_\beta(0) = G_\beta(\beta)$, \ba && G_\beta(\tau_1- \tau_2) =\left\langle \tau_1\left|\frac{1}{-\partial_\tau^2 + \omega^2}\right|\tau_2\right\rangle \no && = \frac{1}{2\omega\sinh \frac{\omega\beta}{2}}\cosh\left\{\omega\left(\frac{\beta}{2}-|\tau_1- \tau_2|\right)\right\} .\ea Our task is to perform the local change of the variable $\sigma(\tau)$ so that to find explicitly   this functional \gl{varharm} in terms of new variables.

Let's introduce  the nonlinear transformation, \be \tilde\omega^2 (\tau) = \omega^2 + i \sigma(\tau) = -\dot u(\tau) + u^2(\tau) , \la{atrans} \ee with $u(\tau)$ being periodic, $u(0)= u(\beta)$ . One can show \cite{AAA} that in order to provide the one-to-one correspondence between $\sigma (\tau)$ and $u (\tau)$ one has to fix the sign of Re $u(\tau)$, for a definiteness, we adopt Re $u(\tau) > 0$.

The original action can be re-expressed as \be S({\vec q},u) = \int\limits_{0}^{\beta} d\tau \left[{\dot{\vec q} }(\tau) + u(\tau){\vec q }(\tau)\right]^2 , \ee after exploiting the periodicity of involved functions.

Consequently  the calculation of the functional determinant \gl{varharm} is equivalent to evaluation of the Jacobian of the following change of (periodic) variables, \ba &&{\dot{\vec q} }(\tau) + u(\tau){\vec q }(\tau)= {\dot{\vec q}_1 }(\tau) +  \omega{\vec q}_1(\tau);\la{change} \\ && {\vec q}_1(\tau) = {\vec q}(\tau) + \int\limits_0^\beta G^{(1)}_\beta(\tau- \tau_1)\ [u(\tau_1)-\omega]\ {\vec q}(\tau_1)\ d\tau_1 , \nonumber \ea where the periodic Green function for inverse first-order differential operator  is given by \ba G^{(1)}_\beta(\tau- \tau_1) &=& \exp[-\omega(\tau- \tau_1)]\left\{\theta (\tau- \tau_1)\right.\no &&\left.+\frac{1}{\exp(\omega\beta)-1}\right\} ,\la{firstor} \ea with $\theta(x)$ being a step function. Eventually in these variables the Jacobian reads, \be \left|\!\left| \frac{\delta {\vec q }_1 }{\delta {\vec q }}\right|\!\right| = \left[ \frac{\sinh \frac12 \int\limits_0^\beta d\tau u(\tau)}{\sinh \frac12 \omega\beta} \right]^N , \la{jac1} \ee and therefore the partition function becomes a local function of $\int\limits_0^\beta d\tau u(\tau)$, \ba {\cal Z}_\beta(u) &=& \left(2\sinh \frac{\omega\beta}{2}\right)^{- N} \left|\!\left| \frac{\delta {\vec q } }{\delta {\vec q }_1}\right|\!\right|\no &=&  \left[2\sinh \int\limits_0^\beta d\tau \frac12 u(\tau) \right]^{- N}. \la{harm2} \ea 
This is the master formula of the EO-transformation. We remark that in the final expression the parameter $\omega$ is hidden in the definition of the variable $u(\tau)$, namely Re $(-\dot u + u^2) = \omega^2$, Im $(-\dot u + u^2)$ is arbitrary,  according to Eq. \gl{atrans} . The overall sign of the denominator in \gl{harm2} conforms to the condition Re $u >0$ . For the opposite choice Re $u < 0$ the sign of the integral in \gl{harm2} must be changed $u \rightarrow - u$ . One can use Eq. \gl{harm2} directly to find the partition function for a temperature dependent oscillator frequency, just fitting $u(\tau)$ instead of $\sigma(\tau)$.

However we intend to use the functional of $\sigma(\tau)$ to perform the Legendre transformation from the multidimensional $O(N)$ symmetric QM to a dual one-dimensional QM with the same energy spectrum. Namely let us consider the  spherically symmetric oscillator potential $\omega^2{\vec q \ }^2 + V_{an}({\vec q \ }^2)$  with an anharmonicity $V_{an}$ bounded from below and substitute the quadratic invariant ${\vec q \ }^2 (\tau)$ by an independent variable $\gamma(\tau)$ . In the path integral for partition function it can be realized by means of the auxiliary Lagrange multiplier $\sigma(\tau)$, \ba \int \frac{{\cal D} \sigma(\tau)}{const} \exp\left\{i \int\limits_0^\beta d\tau \sigma(\tau)\Big( \gamma(\tau)-{\vec q \ }^2(\tau)\Big)\right\} . \la{dual} \ea Certainly the usage of the function $\sigma(\tau)$ as a new dynamic variable in the path integral \gl{genf} leads to a non-local Quantum Mechanics of this "observable" due to \gl{nonloc} and \gl{varharm}. Meanwhile the  EO-transformation \gl{atrans} makes this  QM local in terms of the complex variable $u(\tau)$ as is encoded in Eq. \gl{harm2}.

To complete the EO-transformation it remains to calculate its Jacobian  in the integral \gl{dual}, \ba &&\delta\sigma(\tau) = i \Big(\partial_\tau -2 u(\tau)\Big) \delta u(\tau);\no &&\left|\!\left| -i\frac{\delta\sigma}{\delta u}\right|\!\right| = \left|\!\left| \Big(\partial_\tau -2 u(\tau)\Big)\right|\!\right| . \la{period} \ea This Jacobian for periodic boundary conditions has the same structure as the Jacobian in the change of coordinates \gl{change} and can be calculated in a similar way, \be \left|\!\left| -i\frac{\delta\sigma}{\delta u}\right|\!\right| = \tilde C \cdot 2\sinh \int\limits_0^\beta d\tau u(\tau). \la{jac2} \ee The infinite constant $\tilde C$ must be absorbed by the redefinition of the path integral normalization. The overall sign in \gl{jac2} again conforms to the condition Re $u >0$ .

We remark that in the limit $u \rightarrow 0$ this Jacobian vanishes as  the relation \gl{period} between $\delta\sigma(\tau)$ and $\delta u(\tau)$ in this limit is not invertible. Instead for Re $u > 0$ it is invertible and the Jacobian \gl{jac2} does not vanish.
\section{Dual potential systems\label{aga}}
Thus the entire duality transformation relates the N-dimensional $O(N)$ symmetric system with a potential $\omega^2 {\vec q \ }^2 + V_{an}({\vec q \ }^2)$ to the one-dimensional quantum system with complex momenta, \ba &&{\cal Z}_\beta =  \int\limits_{\vec{q}(0) = \vec{q}(\beta)} {\cal D}^N q(\tau)\no\times&& \exp \left\{ - \int\limits_{0}^{\beta} d\tau \left[{\dot{\vec q}\ }^2(\tau) + \omega^2{\vec q \ }^2(\tau)+ V_{an}\Big({\vec q \ }^2 (\tau)\Big)\right]\right\}\no =&& \int\limits_{u, \gamma(0) = u, \gamma(\beta)} {\cal D}u(\tau) {\cal D}\gamma(\tau) \exp \Big\{  \int\limits_{0}^{\beta} d\tau \Big[u(\tau)\dot\gamma(\tau)\no +&& \gamma(\tau)u^2(\tau) -\omega^2 \gamma(\tau)- V_{an}\Big(\gamma(\tau)\Big) \Big] \Big\} \Phi \left(\int\limits_0^\beta d\tau u(\tau)\right),\no && \Phi (z) = \frac{2\sinh z}{\left[2\sinh(z/2)\right]^N} ,\la{genpot} \ea where the complex momentum $u$ runs along the trajectories with Re $(-\dot u + u^2) = \omega^2$,  and Re $u > 0$. We notice that although the variable $\gamma(\tau)$ replaces the positive invariant ${\vec q \ }^2(\tau)$ it can be also treated  as an unconstrained real variable running along the entire axis according to \gl{dual} . To make it consistent one can  define $V_{an} \equiv V_{an}\Big(|\gamma(\tau)|\Big) .$

This representation contains the canonically conjugated variables  $\gamma (\tau)$ and $ u (\tau)$  with the kinetic term $\gamma (\tau) u^2 (\tau)$ mixing both conjugated variables. Therefore the path integral suffers from the ordering problem \cite{alim,mizrahi}. As it was pointed out in \cite{AAA} the Weyl ordering suits well to justify the periodicity conditions and the vanishing of boundary contributions.

Expanding the function, \ba &&\Phi (z)  = \sum\limits_{l=0}^\infty a^{(N)}_l  \ \exp\Big(-\big(\frac{N}{2}+l-1\big)z\Big);\no && a^{(N)}_l \equiv (N+2l -2)
\frac{(N+l-3)!}{(N-2)!\ l!}\ \ \mbox{for}\  N > 1; \ \la{exp1}\\
&&\Phi (z)  = \exp(\frac12 z) + \exp (- \frac12 z) \ \ \mbox{for}\  N = 1, \no &&\mbox{\it i.e.}\quad a^{(1)}_0 = a^{(1)}_1 = 1;\ a^{(1)}_l\big|_{l\geq 2} = 0 , \ \la{exp2} \ea one obtains the statistical weights (degeneracy numbers) in the decomposition of the partition function into the sectors with a given angular momentum $l$, \ba &&{\cal Z}_\beta = \sum\limits_{l=0}^\infty a^{(N)}_l {\cal Z}_l (\beta) ;\no
&&{\cal Z}^l_\beta  = \int\limits_{u, \gamma(0) = u, \gamma(\beta)} {\cal D}u(\tau) {\cal D}\gamma(\tau)\exp \left\{- S[u,\gamma] \right\}\no&&= \int\limits_{u, \gamma(0) = u, \gamma(\beta)} {\cal D}u(\tau) {\cal D}\gamma(\tau) \la{partial}\\
&&\times\exp \Big\{ - \int\limits_{0}^{\beta} d\tau \Big[-u(\tau)\dot\gamma(\tau)- \gamma(\tau)u^2(\tau)\no&& +\omega^2 \gamma(\tau)+ V_{an}\Big(\gamma(\tau)\Big) +\Big(\frac{N}{2}+l-1\Big) u(\tau)\Big] \Big\} .\nonumber \ea The weights $a^{(N)}_l $ correctly reproduce the degeneracy numbers for $N>1$, for instance, in the case of $O(2)$ symmetry $a^{(2)}_l = 2$, for $O(3)$ symmetry $a^{(2)}_l = 2l +1$ etc.

The further evaluation may be two-fold following a choice of integration domain for the variable $\gamma(\tau)$. According to the ansatz \gl{dual} one can use either constrained, non-negative $\gamma = \rho^2 \geq 0$ or unconstrained $-\infty < \gamma < \infty$. For the first choice one may safely deform the complex path $u(\tau) = \rho(\tau){\tilde u}(\tau)$ holding Re $ {\tilde u}(\tau) > 0$ and therefore not intersecting the poles of $\Phi \left(\int\limits_0^\beta d\tau u(\tau)\right)$ in the path integral \gl{genpot}. The further shift ${\tilde u}(\tau) \rightarrow {\tilde u}(\tau) - \dot\rho$ leaves the functional $\Phi$ invariant due to periodicity $\rho(0)= \rho(\beta) .$  Finally, for $N\geq 2$ one can complete this evaluation by contour deformation
$${\tilde u}(\tau) \rightarrow {\tilde u}(\tau) +  \frac{N + 2l -2}{2\rho(\tau)}$$
which again does not cross the poles of $\Phi(z)$ just increasing the real part of complex momentum variable. One obtains, \ba &&{\cal Z}_\beta^{l}   = \int\limits_{\rho(0) = \rho(\beta)} {\cal D}\rho(\tau)\ \prod\limits_\tau \theta (\rho(\tau)) \exp \Big\{ - \int\limits_{0}^{\beta} d\tau \Big[\dot\rho^2(\tau)\no&& +\omega^2 \rho^2(\tau)+ V\Big(\rho^2(\tau)\Big) +\frac{\Big(N+2l-2\Big)^2 - 1 }{16\rho^2(\tau)}\Big] \Big\}, \la{partialrho} \ea which is nothing but a partition function for the reduced radial Schr\"odinger equation for $N$-dimensional particle in the spherically symmetric potential for a given angular momentum $l$. The extra contribution $-1/16\rho^2$ in the centrifugal term of \gl{partialrho} is due to correct resolution of ordering problem \cite{AAA} in the chain of above transformations . The representation \gl{partialrho} is justifiable for $N \geq 2$ . However it is not valid for $N< 2$ due to the fact that  the EO-transformation is unambiguously defined only for Re $u > 0$ and for $N< 2$ the shift of $u(\tau)$ required to calculate the Gaussian path integral crosses the line Re $u = 0$ confronting the pole problem in $\Phi (z)$ of Eq.\gl{genpot}.

For the second choice of the unconstrained coordinate variable $\gamma(\tau)$ , in a special case of quartic anharmonicity $V_{an}({\vec q \ }^2)= \lambda({\vec q \ }^2)^2$ , one can instead integrate in  $\gamma(\tau)$ to derive, \ba &&S[u,\gamma; \lambda] \Rightarrow S[u; \lambda] =\int\limits_{0}^{\beta} d\tau \Big[-\frac{(\dot u(\tau))^2}{4\lambda}\no&&- \frac{\big(u^2(\tau) -\omega^2\big)^2}{4\lambda} +\Big(\frac{N}{2}+l-1\Big) u(\tau)\Big] \la{anhar}. \ea This representation is not subject to any bounds on $N$ and is valid also for $N < 2$ .

In perturbation theory one expands \cite{AAA}  around a zero-order stationary configuration along the imaginary axis $u(\tau) = \omega + i2\sqrt{\lambda} v(\tau)$ with a real fluctuation $ v (\tau)$ . The latter is achieved by the  deformation of complex paths $u(\tau)$ , \ba &&S[v; \lambda] =\int\limits_{0}^{\beta} d\tau \Big[(\dot v(\tau))^2- 4\lambda v^4(\tau) + 8i\sqrt{\lambda} \omega v^3(\tau)\no&& + 4 \omega^2 v^2(\tau)  +\Big(\frac{N}{2}+l-1\Big) (\omega + i2\sqrt{\lambda} v(\tau))\Big] \la{anharv}. \ea
 As a result, within the perturbation theory, the set of anharmonic oscillators with the complex quartic potential unbounded from below is discovered to describe the same energy spectrum as the original real $O(N)$ symmetric  Hamiltonian in \gl{genpot} with the potential bounded from below.

This equivalence was first found for the dim-2 oscillator \cite{zinn,avron} in the singlet sector $l=0$ in several (up to 15) orders of perturbation theory (see the discussion in \cite{zinnrev}) and proved for arbitrary $l$ in \cite{AAA}. It was confirmed  later in \cite{busl} using the technique of differential equations and integral transforms. For one-dimensional oscillator $N=1$  with $\omega = 0$ a similar equivalence has been derived recently  in \cite{jones,bender} . In the latter case evidently two branches  of trajectories $u$ with different signs of $\omega$ coalesce. The case $N=1$ needs a special comment because one finds in \gl{partial} two contributions into the partition function with the Lagrangians related by reflection $u \leftrightarrow -u$ \cite{jones+}. One comes to its right interpretation in the perturbation theory, say, for the anharmonic oscillator \gl{anharv}. Indeed, the lowest order in the anharmonicity $\lambda$ is saturated by two harmonic oscillators with {\it double} frequencies $2\omega$ and energy shifts $\pm \omega/2$ . Therefore two spectra $\omega (2n +1 \pm 1/2),\ n =0,1,\ldots ,$ jointly form the entire ladder of levels of the initial oscillator $\omega (n + 1/2)$.
\section{Duality for quantum pendulums}
Let us now use the EO-transformation for the analysis of $\sigma$-models which in QM correspond to quantum pendulums. The simplest one is defined by the constraint \gl{dual} for a fixed $\gamma(\tau) = R^2 (\tau)$ embedded  in the path integral for partition function \gl{genf} . It represents a quantum pendulum with variable radius $R (\tau)$ . The EO-transformation eventually produces the decomposition analogous to \gl{genpot} and \gl{partial}, \ba &&{\cal Z}_\beta ( R) =  \int\limits_{\vec{q}(0) = \vec{q}(\beta)} {\cal D}^N q(\tau)\ \prod\limits_\tau \delta\Big({\vec q\ }^2(\tau) - R^2 (\tau) \Big) \no&& \times\exp \left\{ - \int\limits_{0}^{\beta} d\tau \left[{\dot{\vec q}\ }^2(\tau) \right]\right\}\no &&= \int\limits_{\tilde u(0) = \tilde u(\beta)} {\cal D}\tilde u(\tau) \exp \left\{ - \int\limits_{0}^{\beta} d\tau \ \Big[{\dot R}^2 + R^2  \tilde u^2\Big] \right\}\no && \times \Phi \left(i \int\limits_0^\beta d\tau \tilde u(\tau)\right)= \sum\limits_{l=0}^\infty a^{(N)}_l {\cal Z}^l_\beta ( R) ;\no &&{\cal Z}^l_\beta ( R)  = \prod\limits_{\tau =0}^\beta \frac{R_0}{ R (\tau)}\no&&\times \exp \left\{  - \int\limits_{0}^{\beta} d\tau \left[({\dot R}(\tau))^2  + \frac{\Big(N+2l-2\Big)^2}{16 R^2(\tau)} \right]\right\}\no &&\stackrel{R(\tau) = R_0}{\Longrightarrow} \exp\left\{ - \beta\frac{\Big(N+2l-2\Big)^2}{16R_0^2}\right\} . \la{pend} \ea where the last line is obtained for a conventional rigid rotator with  ultralocal quadratic actions admitting the exact calculation  of the energy spectrum \mbox{$E_l =\big(N+2l-2\big)^2/16R_0^2 $} after the contour deformation $ u \rightarrow i\tilde u$. This spectrum dependence of $l$ has been first found in \cite{charap} for $N=3$ with a different value of the ground state energy.  The latter discrepancy is explained by a different choice of pendulum coordinate variables and correspondingly by a different choice of ordering the momenta and coordinates. Our choice is correctly normalized to get a zero ground state energy for $N=2$ when the quantum pendulum is characterized by the Schr\"odinger Hamiltonian $-R^2 \partial_\phi^2$ in the polar-angle coordinate $0\leq \phi < 2\pi$. In this case $E_l =l^2/4R_0^2 $. In the general case of rotator with temperature (Euclidean time) dependent radius $R_{min} \leq R(\tau) \leq R_{max}$ one could normalize
$$\prod\limits_{\tau =0}^\beta \frac{R_0}{ R (\tau)} = 1 \longrightarrow \int\limits_{0}^{\beta} d\tau \ln\frac{R_0}{ R (\tau)} = 0 .$$
\section{Dual form of correlators}
The partition function contains all the information on the energy spectrum and not much knowledge of wave functions. Therefore it makes sense also to analyze its response to external classical sources $\vec\eta(\tau)$ with modification \be S({\vec q}, {\vec\eta}) = S({\vec q}, 0) - \int\limits_{0}^{\beta} d\tau {\vec q\ }\cdot{\vec\eta\ }(\tau) . \la{source} \ee Conventionally the expansion of the partition function \gl{genpot}  with the action \gl{source} in powers of external source $\eta_j(\tau)$ leads to the correlators of coordinate operators at different Euclidean time, \ba &&\Big\langle \hat q_{j(1)}(\tau_1)\cdots \hat q_{j(2n)}(\tau_{2n}) \Big\rangle \no&&= \frac{1}{{\cal Z}_\beta (0)} \frac{\delta}{\delta\eta_{j(1)}(\tau_1)}\cdots \frac{\delta}{\delta\eta_{j(2n)}(\tau_{2n})}{\cal Z}_\beta (\eta)\Big|_{\eta = 0} . \ea The main ingredient of this generating functional is  given by modification of Eq.\gl{genf}, \ba &&{\cal Z}(\sigma,\eta) = \int \frac{{\cal D}^N q(\tau)}{const} \no&&\times\exp \left\{ - \int\limits_{0}^{\beta} d\tau \left[{\dot{\vec q}\ }^2(\tau) + \tilde\omega^2 (\tau) {\vec q \ }^2(\tau) - {\vec q\ }\cdot {\vec\eta\ }(\tau) \right]\right\} \no&& = C \exp\Big\{\Tr{\log(-\partial_\tau^2 + \tilde\omega^2 )} \no&&+ \frac14 \int\limits_{0}^{\beta} d\tau_1 \int\limits_{0}^{\beta} d\tau_2 \sum\limits_{j=1}^N \eta_j (\tau_1) G^{(2)}_\beta(\tau_1, \tau_2 |\sigma)\eta_j(\tau_2) \Big\} , \la{genf+} \ea where we recall that $\tilde\omega^2 = \omega^2 + i\sigma$ . When applying the EO-transformation $\sigma (\tau) \rightarrow u(\tau)$ one comes to the following form of the Green function, \ba &&G^{(2)}_\beta (\tau_1, \tau_2 |\sigma) = \langle \tau|\frac{1}{-\partial_\tau^2 + \tilde\omega^2 }|\tau_1\rangle \\&& \stackrel{\sigma\rightarrow u}{\rightarrow} \int\limits_{0}^{\beta} d\tau \langle \tau_1|\frac{1}{-\partial_\tau + u(\tau)}|\tau\rangle\langle \tau|\frac{1}{\partial_\tau + u(\tau)}|\tau_2\rangle .\nonumber \ea The manifest form of this Green function is easily obtained with the help of convolution of two first-order Green functions of \gl{firstor} type, \ba &&G^{\pm}_\beta (\tau_1, \tau_2 |u) = \langle \tau_1|\frac{1}{\pm\partial_\tau + u(\tau)}|\tau_2\rangle \no&& =\exp\Big(\mp\int\limits^{\tau_1}_{\tau_2} d\tau u(\tau)\Big)\\&&\times\left\{\theta \big[\pm(\tau_1- \tau_2)\big]+\frac{1}{\exp\big(\int\limits^\beta_0 d\tau u(\tau) \big)-1}\right\} \nonumber \ea Once the full information on energy spectrum and wave functions of the Hamiltonian can be derived from correlators it is sufficient to analyze them at zero temperature, {\it i.e.} at $\beta \rightarrow \infty$ .  For this limit let's symmetrize the integration interval $0\leq \tau\leq \beta \Rightarrow -\beta/2 \leq \tau \leq +\beta/2$ and $u(\tau - \frac12\beta)\rightarrow u(\tau)$ to get the entire axis in the zero-temperature limit. Certainly in this limit only lowest, ground state energy level contributes and the generating functional may be well provided by path integral over coordinate functions vanishing at the infinity, $q(\tau) \rightarrow 0$ when $\tau \rightarrow \pm \infty$ . Obviously Eqs. \gl{genpot} and \gl{partialrho} are saturated  with leading contribution for $l=0$ as Re $u > 0$. The formula for correlators \gl{genf+} also survives with the limiting Green function, \ba &&G^{(2)}_\infty(\tau_1, \tau_2 |u) = \int\limits_{-\infty}^{\infty} d\tau \theta (\tau- \tau_1)\theta (\tau- \tau_2) \no&& \times \exp\Big(-\int\limits^{\infty}_{-\infty} d\tau' u(\tau') J(\tau',\tau,\tau_1,\tau_2)\Big) ,
\\ && J^{(2)}(\tau',\tau,\tau_1,\tau_2) \equiv \theta (\tau- \tau')\big[\theta (\tau'- \tau_1) +\theta (\tau'- \tau_2)\big] .\nonumber
\ea Thus the two-point correlator (propagator) of coordinate operators is given by averaging a ``holonomy'' exponential along the contours connecting two points. These contours are to provide the decreasing of exponentials as Re $u > 0$.

After passing through the EO-transformation technology one comes to the generating functional of correlators for an anharmonic potential, \ba
&&{\cal Z}(\eta, \nu)  =  \int \frac{{\cal D}u(\tau) {\cal D}\gamma(\tau)}{const} \la{infty}\\
&&\times\exp \Big\{ - \int\limits_{-\infty}^{\infty} d\tau \Big[-u(\tau)\dot\gamma(\tau)- \gamma(\tau)u^2(\tau)\no&& +\big(\omega^2 + \nu(\tau)\big)\gamma(\tau)+ V_{an}\Big(\gamma(\tau)\Big) +\Big(\frac{N}{2}-1\Big) u(\tau)\Big]\no&&  +\frac14 \int\limits_{-\infty}^{\infty} d\tau_1 \int\limits_{-\infty}^{\infty} d\tau_2 \sum\limits_{j=1}^N \eta_j (\tau_1) G^{(2)}_\infty(\tau_1, \tau_2 |u)\eta_j(\tau_2) \Big\}, \no&& {\cal Z}(0)  = 1 ,\nonumber \ea where we have introduced also the real part of harmonic frequency $\omega^2 \rightarrow \omega^2 + \nu(\tau)$ as an external source which allow to generate independently the correlators of singlet operators $(\vec q\, )^2 $ (see the beginning of this article). This option will produce a number of identities between local and non-local description of the singlet sector (see below).

From \gl{infty} the $n$-point correlators are described with the help of a specific external source in the Euclidean Lagrangian, \ba &&\Big\langle \hat q_{j(1)}(\tau_1)\cdots \hat q_{j(2n)}(\tau_{2n}) \Big\rangle =\frac{1}{2^n}\prod\limits_{l=1}^n\sum\limits_{\{k_l\}\cup\{m_l\}} \no&&\times  \int\limits_{-\infty}^{\infty} d\tau'_l \delta_{j(k_l),j(m_l)} \theta (\tau'_l- \tau_{k_l})\theta (\tau'_l- \tau_{m_l}) \Xi (\tau, \tau') ,\no &&\Xi (\{\tau_k\}, \{\tau'_l\}) \equiv \int \frac{{\cal D}u(\tau){\cal D}\gamma(\tau)}{const} \no &&\times\exp \Big\{ - \int\limits_{-\infty}^{\infty} d\tau \Big[-u(\tau)\dot\gamma(\tau)- \gamma(\tau)u^2(\tau) +\omega^2 \gamma(\tau)\no&&+ V_{an}\Big(\gamma(\tau)\Big) +\Big(\frac{N}{2}-1\Big) u(\tau) + J^{(2n)}(\tau)u(\tau)\Big],\la{dualcor} \ea where $\mbox{dim}\{k_l\}=\mbox{dim}\{m_l\}= n,$ \[  \{k_l\}\cup\{m_l\}= \mbox{Perm}\{1,\cdots,2n\},\ \{k_l\}\cap\{m_l\}= \oslash,\] and the external current includes all contours linking triples of points $\tau'_l$ with $\tau_{k_l}$ and $\tau_{m_l}$, \be J^{(2n)}(\tau) = \sum\limits_{l=1}^n \theta (\tau'_l- \tau)\big[\theta (\tau- \tau_{k_l}) +\theta (\tau- \tau_{m_l})\big]. \ee We remark that at any point $\tau$ this current takes {\it integer} values. In the radial coordinate representation for $ \Xi (\{\tau_k\}, \{\tau'_l\}) $ analogous to \gl{partialrho} one obtains the action, \ba &&S(\rho\big|\{\tau_k\}, \{\tau'_l\} ) = \int\limits_{-\infty}^{\infty} d\tau \Big[\dot\rho^2(\tau) +\omega^2 \rho^2(\tau)\no&&+ V\Big(\rho^2(\tau)\Big) +\frac{\Big(N-2 + 2J^{(2n)}(\tau) \Big)^2 - 1}{16\rho^2(\tau)}\Big] . \ea Thus the current $J^{(2n)}(\tau)$ replaces, in fact, an eigenvalue of angular momentum depending on a configuration of piece-wise contours linking $\tau'_l$ with $\tau_{k_l}$ and $\tau_{m_l}$ .

For the case of quantum pendulum with constant radius \gl{pend} one finds the kernel of $n$-point correlator in the following form, \ba
 &&\Xi (\{\tau_k\}, \{\tau'_l\})\\&& = \exp\left\{ - \int\limits_{-\infty}^{\infty} d\tau \frac{J^{(2n)}(\tau)(N - 2) + \big(J^{(2n)}(\tau)\big)^2}{4R_0^2}\right\} ,\nonumber
\ea which provides the $n$-point correlator \gl{dualcor} with an integral representation corresponding to generalized hypergeometric functions. Thus  according to the common understanding of what is integrable model, $O(N)$ symmetric quantum pendulums are completely integrable in dual variables. We find it promising in establishing ODE/IM relationship between  one-dimensional differential equations and low-dimensional integrable models \cite{ddt2} .

One can easily find a similar action for the anharmonic oscillator \gl{anhar}, \ba
 &&S[u; \lambda] =\int\limits_{-\infty}^{\infty} d\tau \Big[-\frac{(\dot u(\tau))^2}{4\lambda}- \frac{\big(u^2(\tau) -\omega^2 - \nu(\tau)\big)^2}{4\lambda}\no&& +\Big(\frac{N}{2}+J^{(2n)}(\tau)-1 +\frac{\dot\nu(\tau)}{2\lambda}\Big) u(\tau)\Big] ,
\ea supplemented with the external source $\nu(\tau)$ .
 With its help one can find further non-trivial relations between `` loop'' and local representations of
correlators for coordinate operators. For instance, the v.e.v. of the singlet operator $\Big\langle (\vec q\, )^2 (\tau_1) \Big\rangle$  can be obtained in two ways: first by variation in $\nu(\tau)$, \ba &&\Big\langle (\vec q\, )^2 (\tau_1) \Big\rangle_{\vec q} = -\frac{\delta}{\delta\nu(\tau_1)}{\cal Z}(\eta,\nu)\Big|_{\eta=\nu = 0} \no&& = \frac{1}{2\lambda}\Big\langle \dot u(\tau) -u^2(\tau) +\omega^2 + \nu(\tau)\Big\rangle_u , \la{iden1} \ea
 second,  as a two-point correlator at coinciding arguments
$\tau_1 = \tau_2$ \ba &&\Big\langle (\vec q\, )^2 (\tau_1) \Big\rangle_{\vec q}  = \frac{\delta^2}{\delta^2\vec\eta(\tau_1)}{\cal Z}(\eta,\nu)\Big|_{\eta=\nu = 0}\no&& = \Big\langle \frac{N}{2}\int\limits_{\tau_1}^{\infty} d\tau'\exp\left(- 2 \int\limits_{\tau_1}^{\tau'} d\tau u(\tau)\right) \Big\rangle_u .\la{iden2} \ea Thereby the nonlocal ``loop'' integral \gl{iden2} can be identified with a local differential functional \gl{iden1}.

It is not difficult  to establish a similar relation for quantum pendulums checking that $\Big\langle (\vec q\, )^2 (\tau_1) \Big\rangle_{\vec q} = R^2 $ .\\

Let's make few concluding remarks:\\
We have proved the efficiency of the EO-transformation not only in making different dynamical systems equivalent in their spectra but also in providing the dual form of correlators which guarantees the relationship between wave functions. The correlators for dual oscillators have been recently discussed in \cite{jones} and one can compare our results to see the progress in their description .

 We notice also that an analogous method is applicable to transform the
evolution operator for the harmonic oscillator with a time dependent frequency.

\section*{Acknowledgments}
I give my gratitude to Prof. Ruggero Ferrari for multiple fruitful discussions and  to Dr. Andrea Quadri for useful remarks. This work was supported by the Landau Network- Centro Volta - Cariplo Foundation, grant 2006 and partially by Grant RFBR 06-01-00186-a,, by Grant SAB2005-0140, by  Programs RNP 2.1.1.1112 and LSS-5538.2006.2 .
 
\end{document}